\documentclass[twocolumn,showpacs,preprintnumbers,amsmath,amssymb,floatfix,prc,
superscriptaddress]{revtex4}
\usepackage{graphicx}
\usepackage{dcolumn}
\usepackage{bm}

\def\htm{\hat{t}}

\def\beq{\begin{equation}}
\def\eeq{\end{equation}}
\def\be{\begin{eqnarray}}
\def\ee{\end{eqnarray}}

\newcommand{\pom}{\rm I\!P}
\newcommand{\reg}{\rm I\!R}
\newcommand{\f}[2]{\frac{#1}{#2}}
\newcommand{\areg}{\alpha_{_{\rm I\!R}}}
\newcommand{\apom}{\alpha_{_{\rm I\!P}}}
\newcommand{\mx}{M_{_X}}

\newcommand{\dd}  { {\textrm d}}

\newcommand{\lsim}{
 \mathrel{\setbox0=\hbox{$<$}\raise0.6ex\copy0\kern-\wd0
 \lower0.65ex\hbox{$\sim$}}}

\newcommand{\gsim}{
 \mathrel{\setbox0=\hbox{$>$}\raise0.6ex\copy0\kern-\wd0
 \lower0.65ex\hbox{$\sim$}}}
 
\newcommand{\T}[1]{{\mathrm{#1}}}

\begin{document}
\title{Impact parameter profiles from nuclear shadowing ratio and 
applications to deuteron-gold collisions}
\author{Adeola Adeluyi}
\affiliation{Department of Physics and Astronomy \\
Texas A\&M University-Commerce, Commerce, TX 75429, USA}
\author{Trang Nguyen}
\affiliation{Center for Nuclear Research, Department of Physics \\
Kent State University, Kent, OH 44242, USA}
\author{Bao-An Li}
\affiliation{Department of Physics and Astronomy \\
Texas A\&M University-Commerce, Commerce, TX 75429, USA}

\date{\today}
\begin{abstract}
While current nuclear parton distribution functions (nPDFs) from global fits to 
experimental data are spatially homogeneous, many experimental
observables in nucleus-nucleus collisions are presented in terms of 
centrality cuts. These cuts can be related to impact parameter using 
the Glauber theory and it is thus usual in the description of 
such observables to convolute an assumed impact parameter distribution
with the homogeneous nPDFs. 
In this study we use the Gribov theory of nuclear shadowing supplemented with 
information from diffraction to model the impact parameter distributions of 
nuclear shadowing ratio in the small-$x$ region. The modeled distributions are 
applied to the description of the centrality dependence of observables
in deuteron-gold (d+Au) collisions at $\sqrt{s_{NN}} = 200$ AGeV.
\end{abstract}
\maketitle
\vspace{1cm}
%
%
\section{Introduction}

Nuclear parton distribution functions (nPDFs) encode the modifications to 
free nucleon parton distributions due to the complex, many-body
environment in the nucleus. They are essential in the application of 
perturbative Quantum Chromodynamics (pQCD) to the  
description of many relevant observables in relativistic nucleus-nucleus 
collisions. Most of the currently available nPDFs 
\cite{Eskola:2009uj,Eskola:2008ca,Eskola:1998df,Shad_HKN,
de Florian:2003qf} have been obtained from global fits to diverse
experimental data. These nPDFs from global
fits are homogeneous (independent of impact parameter), and are thus
functions of the Bjorken variable $x$ and squared momentum transferred
$Q^2$. An exception is \cite{Frankfurt:2003zd} which is
based on the Gribov theory \cite{Gribov:1968gs} and  available 
in both homogeneous and inhomogeneous forms.
  
Experimental data on nucleus-nucleus collisions are often presented in terms of 
centrality classes. These classes can be mapped into corresponding intervals 
in impact parameter space using the Glauber formalism. In order to adequately 
describe these data, inhomogeneous (impact-parameter-dependent) parton
distributions are required. Due to the inherent constraints 
in the determination of nPDFs, especially in the gluon sector, it is 
pragmatic not to be limited in the choice of nPDF sets. Thus in order 
to take advantage of the different sets available, there is need for 
phenomenologically robust and realistic prescriptions for generating 
inhomogeneous nPDFs. This is usually carried out by convolution of
impact parameter distributions with homogeneous nPDFs.
One such prescription is given in \cite{Vogt:2004hf} where two
variants of impact parameter distributions are considered. 
The first variant assumes the distribution is proportional to 
the local nuclear density (generally taken as a Wood-Saxon nuclear density). 
The second takes the distribution as proportional to the nuclear path
length (Glauber thickness function). A simpler distribution is used in 
\cite{Li:2001xa} where a hard sphere dependence is assumed. 

Our objective in this study is to model impact parameter distributions 
which can be used with the homogeneous nPDFs in the spirit of
\cite{Vogt:2004hf,Adeluyi:2007qk}. The general idea is to determine
the impact parameter distributions which give a good description of
the nuclear shadowing ratio in the region of small Bjorken $x$. These 
distributions are then applied in conjunction with a homogeneous nPDF set
to calculate the centrality dependence of some observables in 
deuteron-gold collisions. A major simplification underlying the
application is the assumption that the distributions derived from 
small-$x$ shadowing are applicable to modifications at all values of $x$.
As in \cite{Frankfurt:2003zd,Adeluyi:2006xy,Adeluyi:2007wk} we employ a 
generalized form of the Gribov theory, incorporating the real part of the 
diffractive scattering amplitude. Experimental data for nuclear shadowing at 
small $x$ are available from the NMC\cite{Amaudruz:1995tq,Arneodo:1995cs} and
E665\cite{Adams:1992nf,Adams:1995is} experiments. These experimental data are
all at small $Q^{2}$: thus the use of information from diffractive scattering
of real photons ($Q^{2} = 0$ GeV$^2$ at FNAL \cite{Chapin:1985mf}) and 
quasi-real photons ($Q^2 < 0.01$ GeV$^2$ at HERA \cite{Adloff:1997mi}) 
affords a good approximation. 

In order to present a reasonably self-contained analysis, we include all the
relevant details from our previous treatments. The paper is thus organized as 
follows: in Sec.~\ref{shad} we review the basic formalism of Gribov theory as 
applied to shadowing in the small Bjorken-$x$ regime. This section is
essentially the same as in \cite{Adeluyi:2007wk}. We present the results of
our calculations for the shadowing ratio and the impact parameter distributions in 
Sec.~\ref{res}. In Sec.~\ref{App} we illustrate a typical 
application of our results by using the distributions to describe the 
centrality dependence of the nuclear modification factor,
pseudorapidity asymmetry, and hadron ratio in
ultrarelativistic deuteron-gold (d+Au) collisions. We conclude in Sec.~\ref{concl}.

\section{Nuclear Shadowing from Generalized Gribov Theory}
\label{shad}

\subsection{Nuclear Shadowing Ratio}
\label{shadrat}

In nuclei, for small values of the Bjorken variable $x$
($x \lesssim 0.1$), the nuclear structure functions  $F_{2}^{ A}$ are significantly 
reduced compared to the product of the mass number $A$ and the free nucleon
structure function $F_{2}^{ N}$. Since the virtual photon-nucleus cross
section is proportional to $F_{2}$, then equivalently the virtual
photon-nucleus cross section is less than $A$ times the one for free nucleons, 
$\sigma_{\gamma^*  A} < A \,\sigma_{\gamma^*  N}$.
This phenomenon is generally known as nuclear shadowing in the strict sense. 
Similar behavior is observed for real photons at sufficiently high energies 
($\nu \gtrsim 3\,\rm{GeV}$). Thus the nuclear shadowing ratio, 
defined as $F_{2}^{ A}/(A*F_{2}^{ N})$ or alternatively as 
$\sigma_{\gamma^* A}/(A*\sigma_{\gamma^* N})$, is less than unity. 

The (virtual) photon-nucleus cross section is separable into a part which accounts 
for the incoherent scattering from individual nucleons, and a correction 
(shadowing correction) from the coherent interaction with several nucleons:
\begin{equation}
\sigma_{\gamma^*{ A}} = 
Z \,\sigma_{\gamma^* { p}} + (A-Z) \,\sigma_{\gamma^* { n}} 
+ \delta \sigma_{\gamma^* { A}}  \,\,\,  
\end{equation}
The single scattering part is the incoherent sum of 
photon-nucleon cross sections, where $Z$ is the nuclear charge number, 
and $\sigma_{\gamma^* { p}}$ and $\sigma_{\gamma^* { n}}$ are
the photon-proton and photon-neutron cross sections, respectively.
The multiple scattering correction is expressible as an expansion in 
the number of nucleons in the target involved in the coherent scattering 
($n \geq 2$). The dominant contribution to nuclear shadowing comes from 
double scattering, since the probability that the propagating hadronic excitation 
coherently interacts with several nucleons decreases with the number of nucleons. 

We define the shadowing ratio as 
\beq
{\mathcal R}^{S}_{A} = \frac{Z \,\sigma_{\gamma^*{p}}+(A-Z)\,\sigma_{\gamma^*{n}} 
+\delta\sigma_{\gamma^*{A}}}{Z \,\sigma_{\gamma^*{p}}+(A-Z)\,\sigma_{\gamma^*{n}}} \,\,\,  
\eeq 
We neglect the small difference between the photon-proton cross section 
$\sigma_{\gamma {p}}$ and the photon-neutron cross section $\sigma_{\gamma {n}}$ 
in this study, and thus denote by $\sigma_{\gamma^* {N}}$ the generic photon-nucleon 
cross section. The shadowing ratio can thus be written as
\beq \label{eq:sr}
{\mathcal R}^{S}_{A} = 1
+\frac{\delta\sigma_{\gamma^*{A}}}
{A \,\sigma_{\gamma^*{N}}} \,\,\,  
\eeq 
Thus, the evaluation of the shadowing correction, 
$\delta \sigma_{\gamma^* { A}}$, is central to the
calculation of the shadowing ratio. In the next section we utilize the Gribov
theory in a generalized form to determine $\delta \sigma_{\gamma^* { A}}$. 

\subsection{Shadowing Correction From Generalized Gribov Theory}
\label{shadcor}

The original formulation of Gribov is generalized by including the real part 
of the diffractive scattering amplitude. We denote by $\eta$ the 
ratio of the real to imaginary parts of the diffractive scattering amplitude.
In this generalized form the shadowing correction at the level of double 
scattering is given by 
\begin{eqnarray} \label{eq:ds_A}
\delta \sigma_{\gamma^* { A}}=\frac{A(A-1)}{2A^2} \, 16 \pi \, {\cal R}e 
\Bigg[\frac{(1-i\eta)^2}{1+\eta^2} \nonumber\\
\int d^2 b 
\int^{\infty}_{-\infty} dz_1 
\int^{\infty}_{z_1} dz_2 \int_{4 m_{\pi}^2}^{W^2} dM_{ X}^2  
\left. \frac{d^2\sigma^{{\rm diff}}_{\gamma^{*} { N}}}{dM_{ X}^2 dt}
\right|_{t\approx 0} \nonumber \\ 
\rho_{ A}^{(2)}(\vec b,z_1;\vec b, z_2) \, \, 
\exp{\left\{i\frac{(z_1-z_2)}{\lambda}\right\}} \Bigg] \, 
\end{eqnarray} 
\begin{figure}
\begin{center} 
\includegraphics[width=4.5cm, height=4.0cm]{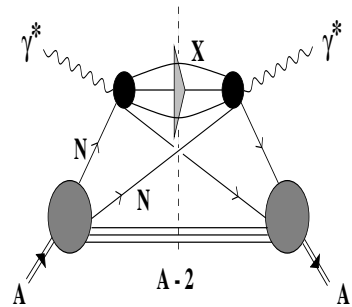}
\end{center}
\caption[...]{Double scattering contribution to $\delta \sigma_{\gamma^* { A}}$.}
\label{fig:double_scattering}
\end{figure}
with $\sigma^{{\rm diff}}_{\gamma^{*} { N}}$ the photon-nucleon 
diffractive cross section. The coherence length, $\lambda$, is given by 
$\lambda = 2\nu/M_{ X}^2$ for real photons. 
As illustrated in Fig. \ref{fig:double_scattering}, 
a diffractive state with invariant mass $M_{ X}$ is 
produced in the interaction of the photon with  a nucleon 
located at position $(\vec b, z_1)$ in the target. 
The hadronic excitation is assumed to propagate 
at fixed impact parameter $\vec b$ and to interact with a second nucleon at
$z_2$.
The probability to find two nucleons in the target 
at the same impact parameter is described by the two-body density 
$\rho_{ A}^{(2)}(\vec b,z_1;\vec b,z_2)$ normalized as 
$\int d^3 r\,d^3 r'\,\rho_{ A}^{(2)}(\vec r, \vec r\,') = A^2$.
The phase factor, $\exp{\{i[(z_1 - z_2)/\lambda]\}}$ in 
Eq.~(\ref{eq:ds_A}) implies that only diffractively  
excited hadrons with a longitudinal propagation length larger than 
the average nucleon-nucleon distance in the target, 
$\lambda > d \simeq 2\,\rm{fm}$, 
can contribute significantly to double scattering.
The limits of integration define the kinematically permitted range 
of diffractive excitations, with their invariant mass $M_{\T X}$ 
above the two-pion production threshold and limited by the 
center-of-mass energy $W=\sqrt{s}$ of the scattering process.

We approximate the two-body density $\rho_{ A}^{(2)}(\vec b,z_1;\vec b,z_2)$
by a product of one-body densities $\rho_{ A}(\vec r) \rho_{ A}(\vec r\,')$
since short-range nucleon-nucleon
correlations are relevant in nuclei only when $z_2-z_1$ is comparable to 
the range of the short-range repulsive part of the nucleon-nucleon force,
i.e. for distances $\lsim 0.4\,\rm{fm}$. 
However, shadowing is negligible in this case and therefore 
short-range correlations are not important in the shadowing 
domain. 

With increasing photon energies or decreasing $x$ down to $x\ll 0.1$, 
the longitudinal propagation length of diffractively excited 
hadrons rises and eventually reaches nuclear dimensions. 
Thus for heavy nuclei, interactions of the excited hadronic state with 
several nucleons in the target become important and should be accounted for. 
Following \cite{Frankfurt:2003zd} we introduce an attenuation factor with an 
effective hadron-nucleon cross section, $\sigma_{{\rm eff}}$. 
The shadowing correction can thus be written as 
\begin{eqnarray} 
\label{eq:ms_A}
\delta \sigma_{\gamma^* { A}}=\frac{A(A-1)}{2A^2} \, 16 \pi \, {\cal R}e 
\Bigg[\frac{(1-i\eta)^2}{1+\eta^2} \nonumber \\ 
\int d^2 b \int^{\infty}_{-\infty} dz_1
\int^{\infty}_{z_1} dz_2 \int_{4 m_{\pi}^2}^{W^2} dM_{ X}^2 
\left. \frac{d^2\sigma^{{\rm diff}}_{\gamma^{*} { N}}}{dM_{ X}^2 dt}
\right|_{t\approx 0} \nonumber\\
\rho_{ A}^{(2)}(\vec b,z_1;\vec b, z_2) \, 
\exp{\left\{i\frac{(z_1-z_2)}{\lambda}\right\}} \nonumber\\ 
\exp{\left\{-(1/2)(1-i\eta)\sigma_{{\rm eff}} \int_{z_1}^{z_2} dz \rho_A(b,z)\right\}} 
\Bigg] \, 
\end{eqnarray}
The effective hadron-nucleon cross section, $\sigma_{{\rm eff}}$ in 
Eq.~(\ref{eq:ms_A}) is defined as
\begin{equation}
\sigma_{{\rm eff}}=\frac{16 \pi}{\sigma_{\gamma {N}}(1+\eta^2)}
\int_{4 m_{\pi}^2}^{W^2} dM_{ X}^2 \left. \frac{d^2\sigma^{{\rm
diff}}_{\gamma^{*} { N}}}{dM_{ X}^2 dt} \right|_{t\approx 0} \,
\label{eq:sigeff} 
\end{equation}
where $\sigma_{\gamma N}$ is the photon-nucleon cross section.
The details of this approach and the approximations inherent in the 
definition of $\sigma_{{\rm eff}}$ are treated thoroughly in
\cite{Frankfurt:2003zd}. For vector mesons as the intermediate hadronic 
excitations, we take $\sigma_{VN}$ as $\sigma_{\rm eff}$  in the attenuation 
factor in Eq.~(\ref{eq:ms_A}), where $\sigma_{V N}$ is the vector meson-nucleon 
scattering cross section.

\subsection{Impact Parameter Profile}
\label{impp}

The shadowing correction (\ref{eq:ms_A}) can be expressed as 
\beq \label{eq:sr2}
\delta \sigma_{\gamma^* { A}} = 
\int\!\! \dd^2b \ {\mathcal F}^{S}_{A}(b) \,\,\,,  
\eeq 
where
\begin{eqnarray} 
\label{eq:ms_A2}
{\mathcal F}^{S}_{A}(b) = 
\frac{(A-1)}{A^2 \, \sigma_{\gamma^*{N}}} \, 8 \pi \, {\cal R}e 
\Bigg[\frac{(1-i\eta)^2}{1+\eta^2} \nonumber \\ 
 \int^{\infty}_{-\infty} dz_1
\int^{\infty}_{z_1} dz_2 \int_{4 m_{\pi}^2}^{W^2} dM_{ X}^2 
\left. \frac{d^2\sigma^{{\rm diff}}_{\gamma^{*} { N}}}{dM_{ X}^2 dt}
\right|_{t\approx 0} \nonumber\\
\rho_{ A}^{(2)}(\vec b,z_1;\vec b, z_2) \, 
\exp{\left\{i\frac{(z_1-z_2)}{\lambda}\right\}} \nonumber\\ 
\exp{\left\{-(1/2)(1-i\eta)\sigma_{{\rm eff}} \int_{z_1}^{z_2} dz \rho_A(b,z)\right\}} 
\Bigg] \, 
\end{eqnarray}
We make use of the fact that the shadowing correction,
$\delta \sigma_{\gamma^* { A}}$,  is a real number and write
\beq \label{eq:sr4}
\int\!\! db \ \frac{ 2\pi b\, {\mathcal F}^{S}_{A}(b)}{{\delta
    \sigma_{\gamma^* { A}}}} = \int\!\! db \ {\mathcal
  G}^{S}_{A}(b) = \ 1\,\,\,,
\eeq 
which is reminiscent of the thickness function distribution:
\beq \label{eq:srr4}
\int\!\! db \ \frac{2\pi b \int\!\! \dd z \rho_{A}({\vec b},z)}
{A}  = \ 1\,\,\,.
\eeq 
The integrands in Eq.~(\ref{eq:sr4}) and Eq.~(\ref{eq:srr4}) for both 
light and heavy nuclei are shown in Sec.~\ref{res}.

\subsection{Diffractive Dissociation}
\label{diffprod}

The shadowing correction, $\delta \sigma_{\gamma^* { A}}$, which is the major
component in the evaluation of the shadowing ratio and impact parameter 
profile, involves the differential diffractive dissociation cross section. The 
determination of this cross section is the subject of this section.

The analysis by the H1 Collaboration\cite{Adloff:1997mi} divides the 
HERA diffractive dissociation data into effectively three  
intervals in $M_{X}^2$: 
the first interval ($0.16 - 1.58$) GeV$^2$ contains
the region of the low-mass vector mesons ($\rho,\,\omega$ and $\phi$), 
the second interval ($1.58 - 4.0$ GeV$^2$) covers the $\rho^{\prime}$ 
resonance region, while the third interval
($M_{ X}^2 > 4.0$ GeV$^{2}$) is that of the high-mass continuum. 
Due to the similarity in the theoretical description of the first and
second intervals, we combine them together.
The differential diffractive cross section is thus written as a 
sum over contributions from two mass intervals,
\begin{eqnarray} 
\label{eq:pDD}
\left. \frac{d \sigma^{D}_{\gamma N}}{d M_X^2 dt} \right|_{t\approx 0} 
= 
\sum_{V=\rho,\omega,\phi,\rho^{'}}\left. 
\frac{d \sigma^{V}_{\gamma N}}{d M_X^2 dt} 
\right|_{t\approx 0} 
+ 
\left.\frac{d \sigma^{cont}_{\gamma N}}{d M_X^2 dt} 
\right|_{t\approx 0} 
\end{eqnarray}
In the following, we briefly summarize the various approximations applied
in these intervals.

\subsubsection{Low-mass vector mesons and the $\rho^{\prime}$ resonances} 
\label{low-mass}

We utilize the generalized vector meson dominance (VMD) \cite{Bauer:1977iq} to 
describe the contribution of the low-mass vector mesons to the differential 
diffractive cross section:
\begin{equation} 
\label{eq:vect_mes_DD}
\left. \frac{d \sigma^{V}_{\gamma N}}{d M_X^2 dt} \right|_{t\approx 0}
= \frac{e^2}{16\pi}\frac{\Pi^{V}(M_X^2)}{M_X^2} \sigma_{V N}^2
\end{equation}
with $\Pi^{V}(M_X^2)$ the vector meson part of the photon spectral function 
$\Pi(M_X^2)$, which is given by
\begin{equation}
\Pi(M_X^2) = \frac{1}{12 \pi^2} \frac{\sigma(e^+ e^- \rightarrow hadrons)}
{\sigma(e^+ e^- \rightarrow \mu^+\mu^-)} \,\,\, 
\end{equation}
In Eq. (\ref{eq:vect_mes_DD}), $\sigma_{V N}$ is the vector meson-nucleon 
cross section and $e^2/4\pi = 1/137$ is the fine structure constant. 
The $\omega$ and $\phi$ mesons are narrow and thus well 
approximated by delta functions. Their contribution to the photon spectral 
function can be written as 
\be
\Pi^{V}(M_X^2) = 
\left(\frac{m_{V}}{g_{V}} \right)^2 
\delta(M_X^2 - m_{V}^2) & ; \,\, & V=\omega,\phi \, ,
\ee
where $m_{V}$ and $g_{V}$, ($V=\omega,\phi$) are the mass and the 
coupling constant of the $\omega$ and $\phi$ mesons, respectively.  

The $\rho$-meson, unlike the $\omega$ and $\phi$ mesons, has a 
large width due to its strong coupling to two-pion states. We have followed
the approach in \cite{Piller:1997ny} and taken this into account through
the $\pi^+\pi^-$ part of the photon spectral function:
\begin{eqnarray} 
\label{eq:Pi_rho}
\Pi^{\rho}\!\left(M_{X}^2\right) = \frac{1}{48 \pi^2} 
\Theta\!\left( M_{X}^2 - 4 m_{\pi}^2 \right) 
\left(1- \frac{4 m_{\pi}^2}{M_{X}^2}\right)^{3/2} \nonumber\\ 
\left| F_{\pi}\left(M_{X}^2\right)\right|^2,
\end{eqnarray}
where $m_{\pi}$ is the mass of the pion
and $M_X=M_{\pi\pi}$ is the invariant mass  
of the $\pi^+\pi^-$ pair. The pion form factor, $F_{\pi}$ is taken from 
Ref. \cite{Klingl:1996by}. A full discussion is given in \cite{Piller:1997ny}. 
We compared the result from the delta function approximation to this more exact
calculation and found that taking into account the width of the
$\rho$-meson increases the differential diffractive cross section by
about 10\%.

The $\rho^{\prime}$ resonance region contains the $\rho(1450)$ and
$\rho(1700)$ mesons which were formerly classified as 
the $\rho(1600)$ \cite{Hagiwara:2002fs}. The FNAL data show an
enhancement in this region and 
we treat this enhancement in terms of an average $\rho^{\prime}$ 
resonance, corresponding to the earlier classification of $\rho(1600)$, as done 
in Ref.~\cite{Chapin:1985mf}. We use the available information on the  
$\rho(1600)$ from Ref. \cite{Bauer:1977iq} in a VMD-type calculation to
evaluate the contribution from this region. The average $\rho^{\prime}$
resonance should have a finite width, but encouraged by the fact that
a delta function in the case of the $\rho$ gives a good
approximation to the full-width result, we employ a narrow resonance
approximation for the $\rho(1600)$: 
\begin{equation} 
\label{eq:vect_mes_D2}
\left. \frac{d \sigma^{V}_{\gamma N}}{d M_X^2 dt} \right|_{t\approx 0}
= \frac{e^2}{16\pi}\frac{\Pi^{V}(M_X^2)}{M_X^2} \sigma_{V N}^2
\end{equation}
with
\begin{equation}
\Pi^{V}(M_X^2) = 
\left(\frac{m_{V}}{g_{V}} \right)^2 
\delta(M_X^2 - m_{V}^2)
\end{equation}
and $V=\rho(1600)$.

\subsubsection{High-mass continuum} 
\label{high-mass}

A full treatment of both the FNAL data and the HERA data in this region has
been carried out by the H1 Collaboration in Ref. \cite{Adloff:1997mi}, using 
the triple-Regge model of photon dissociation. Here we consider only the  
aspects relevant to the present study.
There are two diffractive terms in the triple-Regge expansion: the triple-pomeron
($\pom\pom\pom$) and the pomeron-pomeron-reggeon ($\pom\pom\reg$) terms. The subleading
reggeons have the quantum numbers of the $\rho$, $\omega$, $a_2$ and 
$f_2$ mesons and their trajectories are approximately degenerate. They are 
generally referred to as $\rho$, $\omega$, $a$ and $f$ mesons and their isospin,
signature and $C$- and $G$- parities are $\rho(1--+)$, $\omega(0---)$,
$a(1++-)$ and $f(0+++)$. The pomeron has $\pom(0+++)$ and is thus identical to
the $f$ meson, leading to interference. We neglect such interference effect in
the present study.

The differential dissociation cross section can thus be written as
\begin{eqnarray}
\label{eq:t_pom}
  \frac{ {\rm d^2} \sigma}{{\rm d}\mx^2 \, {\rm d}t} =
  \bigg[\frac{G_{\pom \pom \pom}(0)}{\mx^{2{\apom(0)}}} +
  \frac{G_{\pom \pom \reg}(0)}{\mx^{{4 \apom(0)}-{2 \areg(0)}}}\bigg]  \nonumber \\
  \left( W^2 \right)^{2 \apom(0) - 2} \ 
  e^{B (W^2, \mx^2) \: t} \ ,
\end{eqnarray}
where $B(W^2, \mx^2) = 2b_{p \pom}+2\apom^{\prime} \ln(W^2 / \mx^2)$ 
and $b_{p \pom}$ is the proton-pomeron
slope parameter. $\apom^{\prime}$ is the slope of the pomeron
trajectory while 
$\apom$ and $\areg$ are the pomeron and (effective) reggeon intercept respectively.
We use the values in \cite{Adloff:1997mi} for these parameters. 
Note that the value of $\apom(0)$ ($\apom(0) = 1.068 \pm 0.0492$) agrees within error
with the soft pomeron intercept in Ref. \cite{Donnachie:1992ny} ($\apom(0)
\simeq 1.081$). The usual triple-pomeron approximation corresponds to putting 
$G_{\pom \pom \reg}(0) = 0$ in Eq.~(\ref{eq:t_pom}). The contributions from
the subleading reggeons are small at both NMC and E665 energies
\cite{Adeluyi:2007wk}, thus we use 
the triple-pomeron approximation in this study. We use the value of 
$G_{\pom \pom \pom}(0) = 9.0$ from the analysis in \cite{Adeluyi:2007wk}.

\section{Result: Shadowing Ratio and Impact Parameter Profile}
\label{res}

The treatment outlined in the last section is now applied to 
calculate the shadowing ratio and the impact parameter profile. 
The basic expression is Eq.~(\ref{eq:ms_A2}), 
which involves the ratio of the real to imaginary
amplitudes $\eta$, the photon-nucleon cross section $\sigma_{\gamma {N}}$, 
the nuclear density $\rho_{A}$, and
the effective cross section $\sigma_{{\rm eff}}$ in terms of the diffractive
dissociation cross section.

We use the energy-independent $\eta$'s for the vector mesons from
Ref.~\cite{Bauer:1977iq}. For both $\rho$ and $\omega$ mesons, $\eta$ takes 
values between $0$ and $-0.3$. Here we take $\eta_{\rho} = \eta_{\omega} = -0.2$ 
in accordance with Ref.~\cite{Bauer:1977iq}. The results of our calculation are not
very sensitive to the precise values of $\eta_{\rho(\omega)}$. For the $\phi$ meson,
we take $\eta_{\phi} = 0.13$ \cite{Frankfurt:1998vx}. Due to lack of information, 
we take $\eta_{\rho(1600)} = 0$.
For the high-mass continuum, we follow Ref.~\cite{Frankfurt:2003zd} and define 
$\eta_{\pom}$ as
\begin {equation}
\eta_{\pom} = \frac{\pi}{2}\big(\apom (0)-1\big) \,,
\end{equation}
using the result of Gribov and Migdal \cite{Gribov:1968uy}.
  
We use the Donnachie-Landshoff parameterization of 
$\sigma_{\gamma {p}}$\cite{Donnachie:1992ny} as the generic photon-nucleon 
cross section $\sigma_{\gamma {N}}$.
For the nuclear densities three-parameter Fermi ($3pF$) distributions are applied:
\begin{equation}
\rho (r) = \rho_{0} \frac{1 + \omega(r/R_{A})^2}{1 + e^{(r-R_{A})/d}} \,\, ,
\end{equation}
with the parameter values taken from Ref.~\cite{DeJager:1974dg}. For mass
numbers $A \lesssim 20$ a harmonic oscillator (HO) density distribution may be
more appropriate than the $3pF$ distribution. For uniformity, we use
the $3pF$ distributions for the whole mass range in light of the fact that
uncertainties associated with other parameters are at least comparable.

\subsection{Shadowing Ratio At Small $x$}
\label{Sw25}
\begin{figure}[h]
\begin{center} 
\includegraphics[width=9.5cm, height=9.5cm, angle=270]{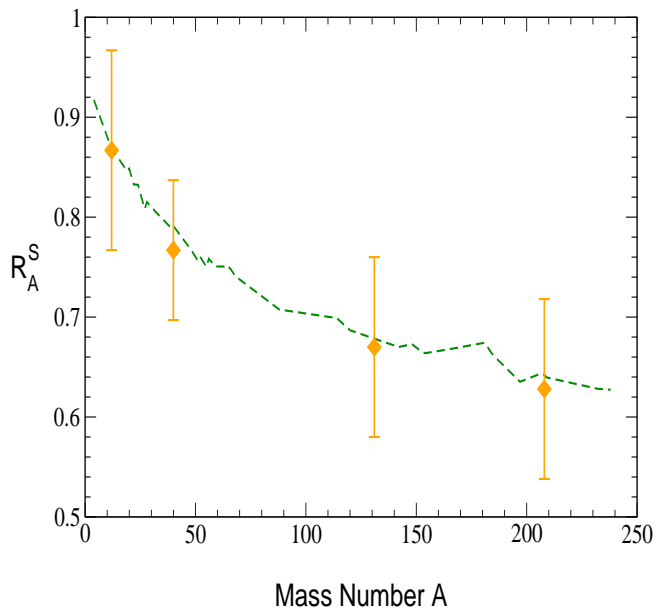}
\end{center}
\caption[...]{(Color Online) Shadowing ratio at small $x$. 
The dashed curve is the result of our calculation.
The shaded triangles are data from the E665 experiment.
}
\label{fig:sr25}
\end{figure}

At very small $x$ ($x\simeq10^{-4}$) the E665 experiment has four data points: 
$^{12}$C, $^{40}$Ca, $^{131}$Xe, and $^{208}$Pb. The results of our calculation
and the experimental data are shown in  Fig.~\ref{fig:sr25}. For small $A$ the
shadowing ratio decreases rapidly with $A$, while for large $A$ the decrease
is more gradual. The result is in good agreement with experiment, and
reproduces quite accurately the trend of the small $x$ shadowing with mass 
number $A$. 

\subsection{Impact Parameter Profile At Small $x$}
\label{impp25}
\begin{figure}[!h]
\begin{center} 
\includegraphics[width=9.5cm, height=9.5cm, angle=270]{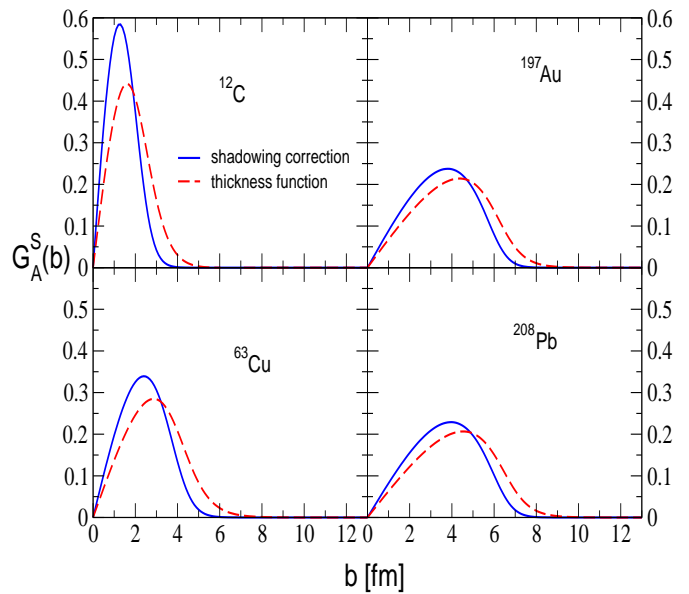}
\end{center}
\caption[...]{(Color Online)  Impact parameter profile at small $x$. The
solid and dashed curves are the results from the present study and 
thickness function dependence respectively.}
\label{fig:spm}
\end{figure}

We use Eq.~(\ref{eq:sr4}) to determine the impact parameter profile for 
nuclei in the mass range $3 < A < 239$ and at small $x$ ($x\simeq 10^{-4}$).
In Fig.~\ref{fig:spm} we present the profile as a function of impact
parameter b for four representative 
nuclei: $^{12}$C, $^{63}$Cu, $^{197}$Au, and $^{208}$Pb. These
results are shown in the figure as solid curves.  

As noted in the introduction, two prescriptions are given in
\cite{Vogt:2004hf}: a local density dependence and a Glauber thickness function
scaling. 
At small Bjorken-$x$ the thickness function dependence is probably
more physical. Its impact parameter distribution, from
Eq.~(\ref{eq:srr4}), is shown as dashed curves in Fig.~\ref{fig:spm}. 
For both the shadowing correction and thickness function, the profiles
have narrow widths and sharp peaks for light nuclei, with the widths
and peaks broadening out as the mass number increases. Since the
profiles are normalized to unity, it is easy to see their differences:
the profiles from the shadowing correction have taller peaks and a
steeper dropoff to zero than those from the thickness function
dependence.

\begin{figure}[!h]
\hspace{-10.0mm}
\includegraphics[width=10.0cm, height=9.5cm, angle=270]{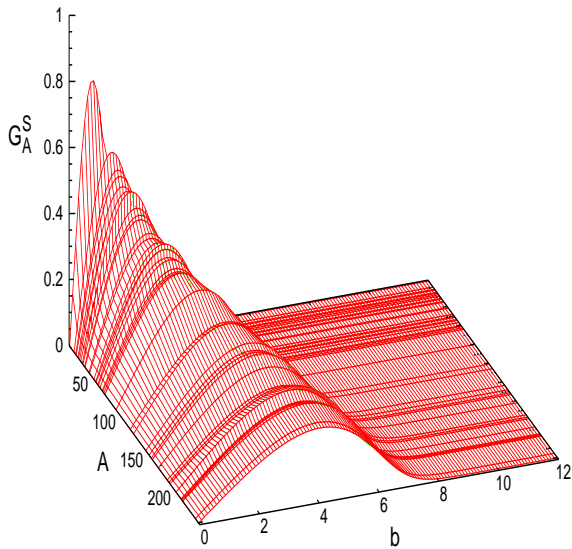}
\caption[...]{(Color On line)  Impact parameter profile at small $x$ as
a function of mass number A and impact parameter b. }
\label{fig:mspm}
\end{figure}

Fig.~\ref{fig:spm} only shows the profiles as a function of b for 
four nuclei. In Fig.~\ref{fig:mspm} we show the full result of our
calculation of the profile from shadowing correction as a function of
both the mass number A and impact parameter b. The features noted in
the preceding paragraph are clearly manifested.

For applications to deuteron-gold collisions (Sec.~\ref{App}) it is
convenient to define the dimensionless form of the impact parameter 
distribution. A suitable form for the shadowing correction (sc) is
\beq \label{eq:srr5}
{\mathcal W}_{A}^{sc}(b) = \frac{{\mathcal F}^{S}_{A}(b)}{{\mathcal F}^{S}_{A}(0)}  
\eeq 
while for the thickness function (tf) we can define
\beq \label{eq:srr6}
{\mathcal W}_{A}^{tf}(b) = \frac{\int\!\! \dd z \rho_{A}({\vec b},z)}
{\int\!\! \dd z \rho_{A}(0,z)} \ .  
\eeq 
These forms are not unique: any suitable value of b can be used
in the denominator to render the profiles dimensionless.

\section{Application to deuteron-gold (d+Au) collisions}
\label{App}

We now apply the impact parameter profile discussed in Sec.~\ref{impp25} to 
calculate the centrality dependence of some observable quantities in 
deuteron-gold collisions: nuclear modification factor for $\pi^0$,
$R_{dAu}^{\pi^0}$, pseudorapidity asymmetry, $Y_{Asy}$, for pions and
protons, and hadron-to-pion ratio. 
\subsection{Collinear factorization}
\label{formalism}
The cross section for the d+Au $\!\to\!$ h+X reaction, 
with respect to pseudorapidity $\eta$ and transverse momentum $p_T$ 
can be written as
\begin{eqnarray} \label{eq:dAu}
\frac{\dd\sigma_{dAu}^{h}}{\dd\eta d^2p_T} =   
\sum_{\!\!abcd}\! \int\!\!\dd^2b \ \dd^2s \ \dd x_a \dd x_b \dd z_c \   
t_{d}({\vec s}) \ t_{Au}({|\vec b- \vec s|}) \nonumber \\
F_{\!a/d}(x_a,Q^2,\!{\vec s},z)\
F_{\!b/Au}(x_b,Q^2,\!{|\vec b- \vec s|},z')\ \nonumber \\
\frac{\dd\sigma(ab\!\to\!cd)}{\dd\htm}\,
\frac{D_{h/c}(z_c,\! Q_{f}^2)}{\pi z_c^2} \hat{s} \, \delta(\hat{s}+\htm+\hat{u}) \,\, ,
\end{eqnarray}
where $x_a$ and $x_b$ are parton momentum fractions in deuteron and gold, respectively, 
and $z_c$ is the fraction of the parton momentum carried by the final-state hadron~$h$.
The factorization and fragmentation scales are $Q$ and $Q_f$ respectively. 
Here
\begin{equation} \label{eq:tf_1}
t_{A}({\vec s}) = \int\!\! \dd z \rho_{A}(\!{\vec s},z)
\end{equation}
is the Glauber thickness function of nucleus $A$, with the nuclear density distribution,
$\rho_{A}({\vec s},z)$ subject to the normalization condition
\begin{equation} \label{eq:tf_2}
\int\!\! \dd^2s \ \dd z \rho_{A}({\vec s},z) = A \,\, .
\end{equation}
The quantity $\dd\sigma(ab\!\to\!cd)/\dd\htm$ in Eq.~(\ref{eq:dAu}) represents the 
perturbatively calculable partonic cross section, and $D_{h/c}(z_c,\!{Q_{f}}^2)$ stands 
for the fragmentation function of parton $c$ to produce hadron $h$, evaluated at momentum 
fraction $z_c$ and fragmentation scale $Q_f$. 

Since we are interested in the description of centrality-selected data
using nPDFs which are spatially homogeneous, we need to incorporate
some form of impact parameter dependency. We follow the
approach in \cite{Vogt:2004hf,Adeluyi:2007qk} and write
\begin{equation} \label{eq:F_1}
F_{\!a/A}(x,Q^2,\!{\vec s},z) = f_{\!a/N}(x,Q^2)S(A,x,Q^2,\!{\vec s},z).
\end{equation}
Here 
$f_{\!a/N}(x,Q^2)$ is the nucleon parton distribution function, which can be expressed as 
\begin{equation} \label{eq:F_2}
f_{\!a/N}(x,Q^2) = 
\f{Z}{A} f_{a/p}(x,Q^2) + (1-\f{Z}{A}) f_{a/n}(x,Q^2) \,\, ,
\end{equation}
with $f_{a/p}(x,Q^2)$ ($f_{a/n}(x,Q^2)$) being the proton (neutron) parton
distribution function as a function of Bjorken $x$ and factorization scale $Q$.
The inhomogeneous shadowing function is defined as
\begin{equation} \label{eq:F_3}
S(A,x,Q^2,\!{\vec s},z) =  
1 + N_{\Phi}[S^{\prime}(A,x,Q^2) - 1]\Phi({\vec s},z)
\end{equation}
with $\Phi({\vec s},z)$ dimensionless and $N_{\Phi}$ a suitable 
normalization constant such that 
\begin{equation} \label{eq:F_4}
\frac{1}{A}\int\!\! d^2s dz \rho_{A}(\!{\vec s},z)S(A,x,Q^2,\!{\vec s},z) = 
S^{\prime}(A,x,Q^2)
\end{equation}
Here $S^{\prime}(A,x,Q^2)$ denotes the homogeneous shadowing function. Note
that the homogeneous nPDFs are the product of the homogeneous shadowing function
$S^{\prime}(A,x,Q^2)$ and the nucleonic PDFs $f_{\!a/N}(x,Q^2)$.
The dimensionless function $\Phi({\vec s},z)$ furnishes the required impact
parameter dependency. We use the profile derived from the shadowing
correction, ${\mathcal W}_{A}^{sc}(b)$ (Eq.~(\ref{eq:srr5})), subject
to the normalization condition in Eq.~(\ref{eq:F_4}).

The density distribution of the deuteron is obtained from the Hulthen 
wave function\cite{Hulthen1957} (as in Ref.~\cite{Kharzeev:2002ei}), while the
density distribution for gold is taken as a Woods-Saxon distribution with
parameters from Ref.~\cite{DeJager:1974dg}. We fix the scales as $Q = Q_f = p_T$, 
where $p_T$ is the transverse momentum of the final hadron. 
The partonic differential cross sections, $\dd\sigma(ab\!\to\!cd)/\dd\htm$ 
were evaluated at leading order (LO). We note that, if a $K$ factor
was used to approximate the effects of higher orders, these effects
would cancel in the ratios calculated in the present study. 
We use the Eskola-Paukkunen-Salgado (EPS08)\cite{Eskola:2008ca} 
shadowing functions with the MRST2001 leading order (LO) parton distribution 
functions (PDFs)\cite{Martin:2001es}. For the
fragmentation functions we use the DSS set \cite{deFlorian:2007aj,
de Florian:2007hc} for pions and protons.

As stated earlier experimental data are often presented in terms of 
centrality classes, the most central class corresponding to head-on
collisions. The Glauber description of heavy ion collisions afford a 
connection between centrality classes and impact parameter cuts. An
analytic expression valid to a very good approximation is given by
\cite{Broniowski:2001ei}
\begin{equation} \label{eq:c_b}
c \approx \frac{\pi b^2}{\sigma_{dAu}^{inel}}
\end{equation}
where c is the centrality, b the impact parameter, and 
$\sigma_{dAu}^{inel}$ is the inelastic deuteron-gold cross section.

\subsection{Nuclear Modification Factor}
\label{modfacts}
The d+Au nuclear modification factor for a hadron h, $R_{dAu}^h$, is defined as 
\begin{equation}
\!R_{dAu}^h(p_{T}) =
 \frac{1}{\langle N_{bin} \rangle} \, \frac{\dd\sigma_{dAu}^{h}}{\dd\eta \ \dd^2p_T} \left/
 \frac{\dd\sigma_{pp}^{h}}{\dd\eta \ d^2p_T} \right. \,\, ,
\label{RdAu}
\end{equation}
where the average number of binary collisions, $\langle N_{bin} \rangle$ 
in the various impact-parameter bins is given by
\begin{equation}
  \langle N_{bin} \rangle = \langle \sigma_{NN}^{inel}\ T_{dAu}(b) \rangle \,\, .
\end{equation}
Here $\sigma_{NN}^{inel}$ is the inelastic nucleon-nucleon 
cross section and 
\beq
T_{dAu}(b) = \int\!\! \dd^2s \ t_{d}({\vec s}) \ t_{Au}({|\vec b- \vec s|})
\eeq
represents the deuteron-gold nuclear overlap function. The nuclear modification 
factor $R_{dAu}^h$ is thus just the ratio of the d+Au and proton-proton (pp) cross 
sections, normalized by the average
number of binary collisions,~$\langle N_{bin} \rangle$. 

\begin{figure}[!h]
\begin{center} 
\includegraphics[width=9.5cm, height=9.5cm, angle=270]{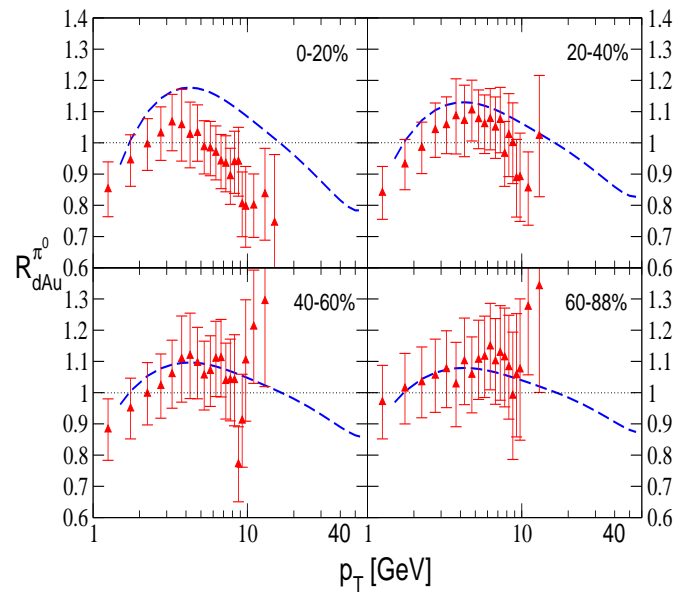}
\end{center}
\caption[...]{(Color Online) Nuclear modification factor $R_{dAu}$ for
  neutral pion in different centrality classes. Filled triangles denote
  experimental data from the PHENIX Collaboration \cite{Adler:2006wg}.}
\label{fig:rdaum0}
\end{figure}

The PHENIX Collaboration has presented results for the nuclear
modification factor for neutral pions \cite{Adler:2006wg} in the pseudorapidity interval
$|\eta| < 0.35$ and four centrality classes: $0-20\%$, $20-40\%$, $40-60\%$,
and $60-88\%$. We use these classes and 
the results of our calculations are displayed in Fig.~\ref{fig:rdaum0}, 
together with the experimental data. 

Our calculation overshoots the data for the
central class ($0-20\%$). This is in part a reflection of the fact
that the minimum bias EPS08 overpredicts the data by about
$5\%$. The data shows the influence of shadowing
($R_{dAu}^{\pi^0} < 1$) for $p_T < 2.5$
GeV, antishadowing ($R_{dAu}^{\pi^0} > 1$) for $2.5 < p_T < 6$ GeV,
and the EMC effect ($R_{dAu}^{\pi^0} < 1$) for $p_T > 6$ GeV. While
our calculation follows this general trend, the data shows a narrower
window for the antishadowing peak relative to calculation. Away from 
the central class, the antishadowing peak progressively broadens with 
increasing centrality, and this is adequately reproduced by calculation.
The agreement with data is in fact quite reasonable for the other
centrality classes under consideration.  
\subsection{Pseudorapidity Asymmetry}
\label{pseudoasym}

As the mechanisms for hadron production in d+Au collisions may be different
at forward rapidities (deuteron side) and backward rapidities (gold side),
it is of interest to study ratios of particle yields between a given 
rapidity value and its negative in these collisions. The STAR 
Collaboration has measured pseudorapidity 
asymmetries, defined as 
\beq
Y_{Asy} = \f{\dd\sigma_{dAu}^{h}}{\dd\eta \ \dd^2p_T}(\mbox{Au-side}) \left/
           \f{\dd\sigma_{dAu}^{h}}{\dd\eta \ \dd^2p_T}(\mbox{d-side}) \right. \,\, ,
\label{yasym}
\eeq
in d+Au collisions for several identified hadron species 
and total charged hadrons in the pseudorapidity intervals 
$|\eta| \le 0.5$ and $0.5 \le |\eta| \le 1.0$ \cite{Abelev:2006pp}. 
Here we consider the pseudorapidity asymmetry for both pions and
protons in different centrality classes for the two STAR
pseudorapidity intervals.

\begin{figure}[!h]
\begin{center} 
\includegraphics[width=9.5cm, height=9.5cm,
  angle=270]{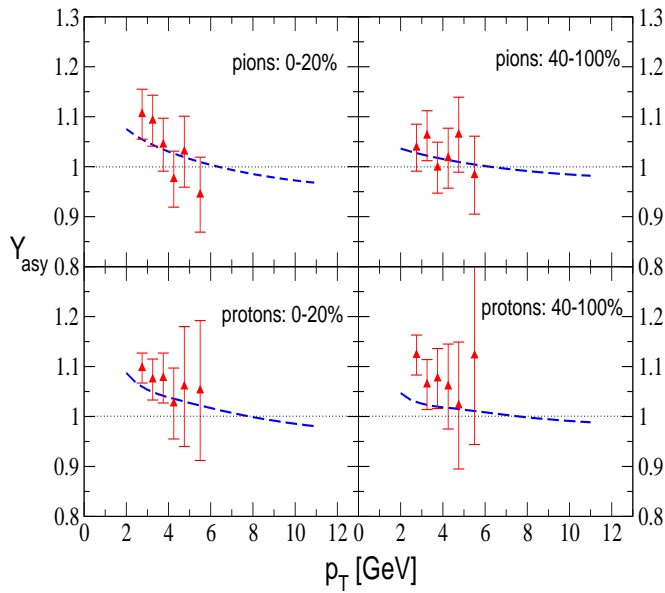}
\end{center}
\caption[...]{(Color Online) 
Centrality dependence of pseudorapidity asymmetry for pions 
(upper panels) and protons (lower panels) in the interval $|\eta| \le
0.5$. Filled triangles denote experimental data from the STAR
Collaboration \cite{Abelev:2006pp}.}
\label{fig:asydg005}
\end{figure}

Fig.~\ref{fig:asydg005} shows the result of our calculation of the
pseudorapidity asymmetry for pions (upper panels) and protons (lower
panels) for the interval $|\eta| \le 0.5$ and centralities $0-20\%$
and $40-100\%$ respectively. 
For the pions, our calculations are in good agreement with data for
both centrality classes, especially for the $40-100\%$ (peripheral)
class where the agreement is manifestly excellent. In the case of
protons, the calculation reproduces the data very well for the
$0-20\%$ (central) class. It is off by a few percent for the peripheral
class. 

Two general remarks are in order: first, the magnitude of the pseudorapidity
asymmetry is rather small in this interval, generally less than or of
the order of $10\%$. This is manifested both by experiment and
calculation, and is due to the fact that the considered interval is
close to midrapidity. Second, the calculation exhibits a greater
degree of asymmetry for the central class relative to the peripheral.

\begin{figure}[!h]
\begin{center} 
\includegraphics[width=9.5cm, height=9.5cm, angle=270]{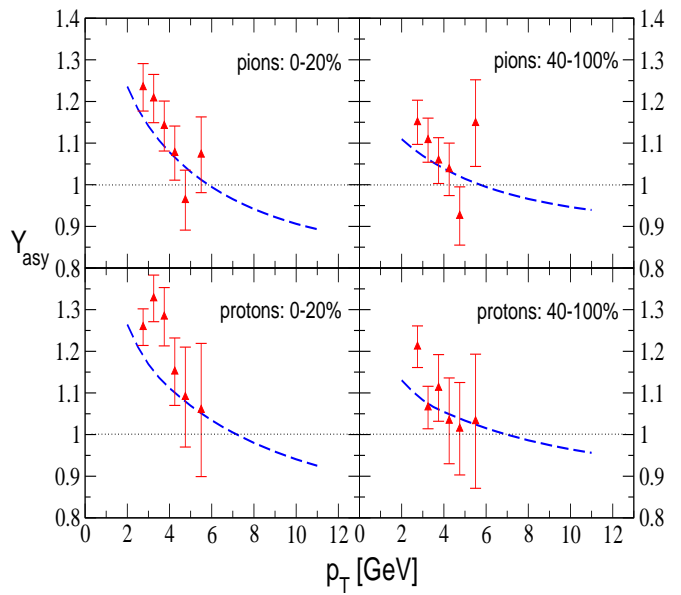}
\end{center}
\caption[...]{(Color Online) 
Centrality dependence of pseudorapidity asymmetry for pions 
(upper panels) and protons (lower panels) in the interval $0.5 \le
|\eta| \le 1.0$. Filled triangles denote experimental data from the
STAR Collaboration \cite{Abelev:2006pp}.}
\label{fig:asydg051}
\end{figure}

The result of our calculation for the pseudorapidity interval 
$0.5 \le |\eta| \le 1.0$ is presented in Fig.~\ref{fig:asydg051}. 
Away from midrapidity, the asymmetry is appreciably enhanced and this
is reflected by both experiment and calculation. Also, as observed in
the previous interval, the magnitude of the asymmetry decreases with 
increasing centrality. Overall, the agreement with data is manifestly 
good for both pions and protons and in both central and peripheral classes.    
%
\subsection{Hadron Ratio}
\label{partrat}
\begin{figure}[!h]
\begin{center} 
\includegraphics[width=9.5cm, height=9.5cm,
  angle=270]{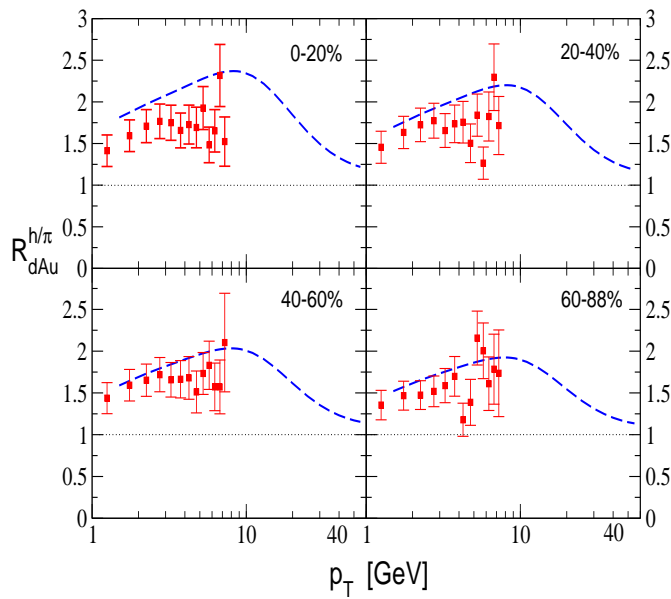}
\end{center}
\caption[...]{(Color Online) 
Centrality dependence of hadron-to-pion ratio. Filled squares denote
  experimental data from the PHENIX Collaboration \cite{:2007by}.}
\label{fig:hadrat}
\end{figure}

As a final example of the application of the impact parameter profile
to d+Au collisions, we consider the centrality dependence of the 
ratio of hadrons to pions, as measured by the PHENIX Collaboration \cite{:2007by}.
We denote this ratio as ${\cal{R}}_{dAu}^{h/\pi}$ and write
\begin{equation}
{\cal{R}}_{dAu}^{h/\pi}(p_{T}) =
 \frac{\dd\sigma_{dAu}^{(h^{+}+h^{-})/2}}{\dd\eta \ \dd^2p_T} \left/
 \frac{\dd\sigma_{dAu}^{\pi^0}}{\dd\eta \ d^2p_T} \right. \,\,.
\label{prat}
\end{equation}
The centrality classes are the same as for the $\pi^0$ nuclear
modification factor, viz: $0-20\%$, $20-40\%$, $40-60\%$, and $60-88\%$. 
Fig.~\ref{fig:hadrat} shows the result of our calculation for these
four centrality classes. The trend also mirrors that observed for 
$R_{dAu}^{\pi^0}$.
Our calculations overshoot the data for the most central class
($0-20\%$) especially at higher $p_T$. The agreement with data is more
reasonable for the other centrality classes.
%
\section{Conclusion}
\label{concl}

We have calculated the the shadowing ratio and impact parameter profiles 
for nuclei in the mass range $3 < A < 240$ and for small $x$ ($x\simeq
10^{-4}$) using a suitably generalized form of the Gribov theory.   
The photon diffractive dissociation cross section, which is the main input 
to our calculation, has been parameterized as a function of the
invariant mass of the diffractively produced hadronic excitation in 
two effective mass intervals. For the low-mass vector mesons and 
$\rho^{\prime}$ resonances, vector meson dominance model is used while 
the high-mass continuum is treated within the framework of
triple-Regge theory. Relevant model parameters are taken from
experiments and earlier studies.
 
The calculated result for the shadowing ratio agrees nicely with small $x$ 
shadowing data from the E665 experiment and also reproduces quite well
the trend of shadowing with mass number $A$. We compare the resultant 
impact parameter profiles for four representative nuclei with the
Glauber thickness function dependence which has been previously applied in
the literature. We also present the profiles as a function of both
mass number A and impact parameter b for $3 < A < 240$ and $0 < b < 12$ fm.

We illustrate the utility of the impact parameter distributions by 
applying them to calculate the centrality dependence of three
different observables in deuteron-gold collisions:
the $\pi^0$ nuclear modification factor $R_{dAu}^{\pi^0}$, 
pseudorapidity asymmetry $Y_{Asy}$ for pions and protons, 
and the hadron-to-pion ratio, ${\cal{R}}_{dAu}^{h/\pi}$. 
Except for the overshoot at the most central class, our calculation
gives an adequate description of the centrality behavior of both
$R_{dAu}^{\pi^0}$ and ${\cal{R}}_{dAu}^{h/\pi}$ as measured by the
PHENIX Collaboration. The agreement with the STAR data for the
centrality dependence of pseudorapidity asymmetry is also good
for both pions and protons. This is encouraging considering the
fact that the distributions have been derived from the small-$x$ 
shadowing ratio, and thus in a strict sense reflects the behavior 
in the shadowing region, i.e $x < 0.1$.
\section{Acknowledgment  s}
\label{ack}
This work was supported in part by the NSF grant PHY0757839.
%
%

\end{document}